\documentstyle[epsfig,galley]{mn}
\def\spose#1{\hbox to 0pt{#1\hss}}
\def\lta{\mathrel{\spose{\lower 3pt\hbox{$\mathchar"218$}}
     \raise 2.0pt\hbox{$\mathchar"13C$}}}
\def\gta{\mathrel{\spose{\lower 3pt\hbox{$\mathchar"218$}}
     \raise 2.0pt\hbox{$\mathchar"13E$}}}
\def\bm{{\beta_M}}
\def\br{{\beta_r}}

\def\del{{\partial}}
\def\alf{{Alfv\'en }}
\def\sL{{\mathcal L}}

\def\grad{{\mathbf \nabla}}
\def\bv{{\mathbf v}}
\def\bb{{\mathbf b}}
\def\bk{{\mathbf k}}

\def\bH{{\mathbf H}}
\def\bB{{\mathbf B}}

\def\dm{{\dot m}}
\def\eps{{\epsilon}}
\def\ez{{\hat{\mathbf z}}}
\def\uk{{\hat{\mathbf k}}}
\def\msun{{\,M_\odot}}
\def\cm{{\rm \,cm}}
\def\K{{\rm \,K}}

\def\gm{{\rm \,g}}

\begin{document}

\title[Photon Bubbles in Discs]{
Photon Bubbles in Accretion Discs
}
\author[C. F. Gammie]
{Charles  F. Gammie$^{1,2,3}$ \\
$^1$ Institute of Astronomy, Madingley Road,
	Cambridge CB3 0HA \\
$^2$ Center for Astrophysics, MS-51, 60 Garden St., 
	Cambridge MA \\
$^3$ Isaac Newton Institute, 20 Clarkson Road,
	Cambridge CB3 0EH
}

\maketitle

\begin{abstract}

We show that radiation dominated accretion discs are likely to suffer from
a ``photon bubble'' instability similar to that described by Arons in
the context of accretion onto neutron star polar caps.  The instability
requires a magnetic field for its existence.  In an asymptotic regime
appropriate to accretion discs, we find that the overstable modes obey
the remarkably simple dispersion relation
$$
\omega^2 = -i g k F(\bB,\bk).
$$
Here $g$ is the vertical gravitational acceleration, $\bB$ the magnetic
field, and $F$ is a geometric factor of order unity that depends on
the relative orientation of the magnetic field and the wavevector.
In the nonlinear outcome it seems likely that the instability will
enhance vertical energy transport and thereby change the structure of
the innermost parts of relativistic accretion discs.

\end{abstract}


\section{Introduction}

Compact objects such as black holes and neutron stars are among the
most interesting objects in astrophysics because of their exotic
strong-field gravitational physics and because they are likely at the
center of some of the most luminous and energetic objects in the
universe.  Since it is mainly their accretion flows, and not the
compact objects themselves, that are readily observed, the dynamics and
radiative properties of the accretion flows are the focus of much
theoretical attention.  

One of the most important models for accretion flows is the thin disc
(Shakura \& Sunyaev 1973, Lynden-Bell \& Pringle 1974).  This model
predicts that for rapidly accreting objects the inner parts of the disc
are radiation pressure dominated.  It was realized soon after the model
was first proposed, however, that radiation pressure dominated discs
are unstable, both viscously (Lightman \& Eardley 1974) and thermally
(Shakura \& Sunyaev 1975).  This suggests that the standard thin disc
model is not self-consistent, and so attention has turned to other
models such as advection-dominated flows (see Narayan 1997 for a
review).

Despite theoretical arguments that the thin disc is unstable, thin disc
spectra are widely used to fit observations of black hole candidates.
Indeed, there is indirect evidence for (theoretically unstable) thin
discs in that some observations are well fit by thin disc spectra.  For
example, some galactic black hole candidates such as Nova Muscae have
X-ray spectra that are well fit by multi-temperature thin disc models
(R. Narayan, private communication).

One can stabilize radiation pressure dominated discs by modifying the
usual prescription for the shear stress $t_{r\phi} \sim \alpha p$ so
that $t_{r\phi} \sim \alpha p_g$ ($p_g \equiv$ gas pressure) or some
combination of gas and radiation pressure (e.g. Lightman \& Eardley
1974, Piran 1978).  Arguments have been advanced in favour of such a
modification by Eardley \& Lightman (1975), Coroniti (1981), Sakimoto
\& Coroniti (1981,1989), and Stella \& Rosner (1984).  These arguments
rely on the thermodynamic peculiarities of magnetic buoyancy in a
radiation dominated plasma.  More recent work has vastly increased our
understanding of magnetically driven angular momentum diffusion in
discs (see the review of Balbus \& Hawley 1997).  Numerical experiments
suggest that $t_{r\phi}$ is limited by Lorentz forces rather than by
buoyant escape of magnetic fields (Stone et al. 1996), although these
experiments do not include radiation pressure and radiative diffusion.

An alternative route to viscous stability (e.g. Liang 1977) is to
modify the disc cooling law using convection, although this is not
successful in eliminating the thermal instability (Piran 1978).
Modification of the disc cooling rate has not seemed a promising
approach.

Recently, however, while investigating neutron star polar cap
accretion, Klein \& Arons (1989, 1991) noticed the development of
evacuated regions, or ``photon bubbles''\footnote{Bubble is something
of a misnomer, since surface tension plays no role.  The phenomenon is
really more like convection.} in their radiation hydrodynamics
simulations.  A subsequent linear analysis (Arons 1992) revealed that
the source of the photon bubbles was an overstable mode present in
radiation dominated, {\it magnetized} atmospheres.  Earlier
incarnations of the photon bubble instability in an unmagnetized plasma
have turned out to be flawed; Marzec (1978) gives a full discussion of
this point.  Later numerical studies of the nonlinear evolution of the
photon bubble instability (Hsu et al. 1997) showed that vertical
transport of energy is enhanced in the nonlinear outcome.

One is naturally led to inquire whether this same photon bubble
instability is present in the radiation dominated parts of accretion
discs, and if so, what the consequences might be for disc structure.
Unfortunately Arons's (1992) analysis is not immediately applicable to
discs.  It considers physical parameters relevant to neutron star polar
cap accretion: a thermal timescale long compared to the dynamical
timescale, negligible gas pressure, and superthermal magnetic field.
Our purpose in this paper is to generalize Arons's work to the regime
appropriate to discs.

First we evaluate conditions inside accretion discs using a standard
$\alpha$ model (\S 2).  Then we write down a set of governing equations
appropriate to these conditions (\S 3).  In \S 4 we construct a model
equilibrium to perturb about, and in \S\S 5 and 6 we work out the linear
theory in the WKB limit.  The astrophysical implications of the result
depend on the nonlinear outcome, about which we speculate in \S 7. \S 8
contains a summary.

\section{Conditions in Accretion Discs}

What conditions are relevant to a study of the photon bubble instability
in accretion discs?  Consider accretion onto a black hole of mass $m
\,\msun$ at a rate $\dot{m} \, L_{edd} c^2$, where $L_{edd} \equiv 4 \pi
G M c/\kappa_{es}$, and $\kappa_{es}$ is the electron scattering opacity.
Using the standard thin-disc, one-zone model (Shakura \& Sunyaev 1973),
assuming radiation pressure is dominant and $\kappa \simeq \kappa_{es}
\simeq 0.4 \gm \cm^{-2}$, we find that at radius $r G M/c^2$
\begin{equation}
\Sigma \simeq 0.37 \,\, \alpha^{-1} \dm^{-1} r^{3/2} \,\, \gm \cm^{-2},
\end{equation}
\begin{equation}
T \simeq 2.4 \times 10^7 \,\, \alpha^{-1/4} m^{-1/4} r^{-3/8}\,\,  \K,
\end{equation}
\begin{equation}
\rho \simeq 4.2 \times 10^{-7}\,\,  \alpha^{-1} m^{-1} \dm^{-2} r^{3/2}\,\,  \gm\cm^{-3}.
\end{equation}
Here and throughout we ignore corrections due to general relativity
and the inner boundary of the disc.  Radiation-dominated accretion
discs have entropy profiles that suggest convective instability
(e.g. Bisnovatyi-Kogan \& Blinnikov 1977).  Calculations of disc vertical
structure using mixing-length theory (Shakura, Sunyaev, \& Zilitinkevich
1978) show, however, that the vertical radiative flux of energy always
dominates the vertical advective flux and so the mean structure of the
disc is not likely to be very different from the values calculated above.

The electron scattering optical depth is
\begin{equation}
\tau_s \simeq \Sigma \kappa_{es}/2 = 0.074 \,\,
	\alpha^{-1} \dm^{-1} r^{3/2},
\end{equation}
while the ``true'' optical depth, which measures the extent to which
radiation is thermalized inside the disc, depends on the absorption opacity
$\kappa_a$:
\begin{equation}
\tau^* = \Sigma \sqrt{\kappa_{es} \kappa_a}/2.
\end{equation}
The absorption opacity is a complicated function of density,
temperature, metallicity and the radiation spectrum.  Taking $\kappa_a
\simeq \kappa_P$, the Planck mean opacity, and evaluating $\kappa_P$
from tables for a solar composition gas (Magee et al. 1995) we find
that $\tau^*$ ranges from $\lta 1$ to $\gg 1$.  For example, at $r =
50$ from a $10 \msun$ black hole accreting at $\dm = 1$, and taking
$\alpha = 0.1$, we find $\tau^* \simeq 14.8$.  Discs around
supermassive black holes have somewhat higher effective optical depths,
since they are cooler and the Planck
mean opacity increases sharply below $3 \times 10^6 \K$ due to
bound-free absorption by metals.  All this implies that the true
optical depth of perturbations that are smaller than the scale height
can be small, and so the radiation field must be treated using an
approximation that is valid in this regime.

The ratio of gas pressure $p_g$ to total pressure $P$ is
\begin{equation}
\beta_r \equiv p_g/P = 1.6 \times 10^{-6}\,\,
	\alpha^{-1/4} m^{-1/4} \dm^{-2} r^{21/8},
\end{equation}
so the boundary of the radiation-dominated region lies at
\begin{equation}
r_c \simeq 160 \,\,
	\alpha^{2/21} m^{2/21} \dm^{16/21}.
\end{equation}
Also the sound speed $c_s$ is
\begin{equation}
c_s/c = 3.0 \,\, \dm r^{-3/2}.
\end{equation}
The radiative diffusivity is $D_0 \equiv c/(\kappa\rho)$.  In
dimensionless form,
\begin{equation}
M_0 \equiv {D_0\over{c_s H}} \simeq 4.5\,\,  \alpha ,
\end{equation}
where $H \equiv$ disc scale height.  Thus $M_0 \sim 1$, while in neutron
star polar cap accretion $M_0 \ll 1$.  This difference changes the
character of the instability significantly, and is the most important
difference between our work and Arons's work.

We also need an estimate for the magnetic field strength.  We use the
prescription $\bm \equiv c_i^2/v_A^2 = (4\alpha)^{-1}$ ($c_i^2 \equiv
P/\rho$ is the isothermal sound speed; $v_A \equiv $ \alf speed),
consistent with the simulations of Hawley, Gammie, \& Balbus (1995).
So if $\alpha = 0.1$, $\bm = 2.5$.  For reasonable values of $\alpha$,
then, $\bm \sim 1$.  Arons (1992) focused on the case $\bm \ll 1$,
but developed a more general analysis in an appendix.

Finally, heat conduction, radiative viscosity, ordinary viscosity,
and ordinary resistivity are all completely negligible.

To summarize the results of this section,  accretion discs have
approximately thermal magnetic fields, so $\bm \equiv c_i^2/v_A^2 \sim
1$.  In their inner region radiation pressure dominates, so $\br 
\ll 1$.  The thermal timescale is comparable to the
dynamical timescale (to within a factor of $\alpha$) so the
dimensionless radiative diffusion rate $M_0 \equiv
(c/(\kappa\rho))/(c_i H) \sim 1$.  Finally, the thermalization length
$l_* \equiv (\rho\sqrt{\kappa \kappa_a})^{-1}$ varies widely but can be
$\sim H$.  Since we will consider perturbations on scales small
compared to the scale height, the perturbations can also have a scale
small compared to the thermalization length.  Thus it is not clear
{\it a priori} that it is appropriate to use the Rosseland diffusion
approximation for the radiation field.

\section{Basic Equations}

Since we cannot use the Rosseland diffusion approximation for the
radiation field, we turn to the more general flux-limited nonequilibrium
diffusion approximation (see Mihalas \& Mihalas 1984 and references
therein).  This approximation is very similar to the Rosseland diffusion
approximation (it still sets $K_{ij} = J \delta_{ij}/3$, where $K$
and $J$ are the second and zeroth angular moments of the intensity),
but, loosely speaking, it allows for the possibility that the radiation
field has a different ``temperature'' than the gas.  More precisely,
it does not require that $J = \sigma T_g^4/\pi$, where $T_g \equiv$
gas temperature.  Because it is a diffusion approximation, it is strictly
valid only on scales large compared to the photon mean free path $(\rho
\kappa_{es})^{-1}$, but it gives qualitatively sensible results on
smaller scales.

The governing equations for the gas, then, are the continuity equation,
\begin{equation}
D_t \rho = -\rho(\grad\cdot\bv),
\end{equation}
where $D_t \equiv \del_t + (\bv\cdot\grad)$, the momentum equation,
\begin{equation}
\rho D_t \bv = -\grad p_g - \rho\grad\phi  +
	{(\bB\cdot\grad)\bB\over{4\pi}} - {\grad B^2\over{8\pi}} 
	+ {4\pi\kappa\rho\over{c}} \bH,
\end{equation}
where $\bH \equiv \int_{d\Omega} I {\bf n}/(4\pi)$ 
\footnote{$H$ denotes the scale height and $H_i$ a component of the
flux.  Likewise $B$ denotes the thermal mean intensity and $B_i$ a
component of the magnetic field.  The difference should be clear in
context.}
is the frequency-integrated flux and $\phi$ is the gravitational
potential, and the gas energy equation ($u \equiv$ internal energy per
unit volume),
\begin{equation}
D_t u = -\gamma u (\grad\cdot\bv) + 4\pi\kappa_a\rho(J - B),
\end{equation}
where $J$ is the frequency-integrated mean intensity, $B \simeq \sigma
T_g^4/\pi$, and we approximate the absorption opacity by the Planck mean
opacity.  The magnetic field evolution is governed by the induction
equation
\begin{equation} D_t \bB = -\bB(\grad\cdot\bv) + (\bB\cdot\grad)\bv
\end{equation} 
and the constraint $\grad \cdot\bB = 0$.  The mean intensity evolution
is given by
\begin{equation} 
{1\over{c}} D_t J - \left({4 J\over{3\rho c}}\right) D_t\rho =
	-\grad\cdot\bH + \kappa_a\rho(B - J),
\end{equation} 
and the flux evolution is given by 
\begin{equation} 
{1\over{c}} D_t \bH = -{1\over{3}}\grad J - \kappa\rho\bH,
\end{equation} 
where we approximate $\kappa$ by the Rosseland mean opacity $\simeq
\kappa_{es}$.

\section{Model Problem: Stratified Atmosphere}

Magnetized, radiation-dominated accretion discs are dynamically
evolving flows, since they are subject to the magnetorotational
instability (Balbus \& Hawley 1991) and, possibly, ordinary convective
instability.  Linear theory therefore cannot even in principle provide
a rigorous guide to their dynamics.  The best we can hope for is to
find a model problem that captures the essence of physical conditions
in accretion discs and is simple enough to solve.

The model problem we have chosen is a nonrotating stratified atmosphere
with $\grad\phi = g \ez = const.$, a {\it uniform} magnetic field,
and a constant flux $H_z$ from below.  This is the least complicated
model that is potentially subject to the photon bubble instability.
It allows us to focus on photon bubbles alone, disentangled from the
magnetorotational instability, magnetic Rayleigh-Taylor instability,
and convective instability.

The model equilibrium is determined by the vertical momentum equation
\begin{equation}
-{1\over{\rho}}{\del p\over{\del z}} + {4\pi\kappa\rho\over{c}} H_z 
	- g = 0,
\end{equation}
the gas energy equation
\begin{equation}
J = B,
\end{equation}
the intensity equation
\begin{equation}
\grad\cdot\bH = 0,
\end{equation}
and the vertical component of the flux equation
\begin{equation}
-{1\over{3}}{d J\over{d z}} = -\kappa\rho H_z.
\end{equation}

This set of equations admits a two-parameter family of solutions.
A natural set of parameters is $\br$ evaluated at $z = z_o$, and $\sL =
H_z 4\pi\kappa/(c g)$, which is the ratio of the flux to the critical flux
where the effective gravity vanishes (i.e., the local Eddington limit).
Defining the pressure scale height $H \equiv c_i^2/g$, at $z = z_o$
we have
\begin{equation}
h_T \equiv - H {\del \ln T_g\over{\del z}} = {1\over{4}}{\sL\over{1 - \br}},
\end{equation}
\begin{equation}
h_\rho \equiv
- H {\del \ln \rho\over{\del z}} = {1\over{\br}}
	\left({1 - \sL{(1 - 3\br/4)\over(1 - \br)}}\right),
\end{equation}
which imply
\begin{equation}
h_r \equiv
- H {\del \ln P_r\over{\del z}} = {\sL\over{1 - \br}},
\end{equation}
\begin{equation}
h_g \equiv
- H {\del \ln P_g\over{\del z}} = {1 - \sL\over{\br}}.
\end{equation}

The entropy profile suggests convective stability (entropy increases 
upwards) if
\begin{equation}
\sL < \sL_{crit} = {4 (\gamma - 1)(4 - 7\br + 3\br^2)
	\over{(\gamma - 1)(16 - 12\br - 3\br^2) + \br^2}}
\end{equation}
(Kutter 1970, Wentzel 1970, Tayler 1954).  In the limit
of small $\br$ this amounts to
\begin{equation}
\sL < 1 - \br + O(\br^2).
\end{equation}
For model atmospheres in which $M_0 \ll 1$ this implies convective
stability.  In discs the question of convective stability is more subtle,
and outside the scope of this paper, because of the presence of radiative
damping and rotation.

\section{Linear Theory}

We will consider only short-wavelength (WKB) perturbations.
Longer-wavelength modes are global in nature and depend on the details
of the equilibrium; but the careful construction of disc equilibria is
a futile endeavour since all weakly magnetized discs are unstable.  The
perturbations have the form
\begin{equation}
\delta f \sim \exp\left[ i (k_x x + \int^z k_z(z') dz' - \omega t)
	\right],
\end{equation}
where $f$ is any one of the perturbed variables.  Like Arons, we take
$k_x = (1 - \mu^2)^{1/2} k$ and $k_z = \mu k$.  The basic small parameter
of the WKB approximation is $(k H)^{-1} \sim \eps \ll 1$ (we assume that
$\mu \sim 1$).

Retaining all terms that are potentially of leading order, the 
linearized continuity equation is
\begin{equation}
-i\omega\delta\rho - \delta v_z {\rho h_\rho\over{H}} = 
	-i \rho \bk\cdot\delta\bv,
\end{equation}
the momentum equation is
\begin{eqnarray}
& -i\omega\rho \delta\bv  = -i \bk \delta p_g
	- i\bk {\bB\cdot\delta\bB \over{4\pi}} 
	+ i \delta \bB {\bk \cdot\bB\over{4\pi}} \hfill \nonumber \\
& \qquad + {4\pi\over{c}}\left(\delta\kappa \rho \bH + \kappa
		\delta\rho\bH + \kappa\rho \delta\bH\right)
	- g \delta\rho\,\ez,
\end{eqnarray}
the induction equation is
\begin{equation}
-i\omega\delta\bB = -i \bB (\bk\cdot\delta\bv) + i(\bk\cdot\bB)\delta\bv,
\end{equation}
the gas energy equation is
\begin{equation}
-i\omega\delta u - \delta v_z {u h_g\over{H}} = 
	- i \gamma u (\bk\cdot\delta\bv) + 4\pi\kappa_a \rho (\delta J -
	\delta B),
\end{equation}
the intensity equation is
\begin{eqnarray}
&{1\over{c}}(-i\omega\delta J - \delta v_z {h_r B\over{H}})
 - {4 J\over{3\rho c}}(-i \rho \bk\cdot\delta\bv) = \hfill \nonumber \\
& \qquad\qquad\qquad - i \bk\cdot\delta\bH + \kappa_a (\delta B - \delta J),
\end{eqnarray}
and the flux equation is
\begin{equation}
{1\over{c}}(i\omega\delta\bH) = - {1\over{3}} i \bk \delta J
	- \kappa \rho\, \delta \bH - \kappa\, \delta\rho\, \bH - \delta\kappa
	\,\rho \bH.
\end{equation}

Together with the constraint $k\cdot\delta\bB = 0$, these equations imply
a complicated eleventh-order dispersion relation $D(\omega,\bk) = 0$ which
we shall not record here (the most compact form of the full relation is
the above equations).  We have confirmed that it contains the dispersion
relations for magnetohydrodynamic (MHD) waves, magnetoatmospheric waves
(e.g. Thomas 1982), internal waves, sound waves in a radiating fluid
(Mihalas \& Mihalas 1984), and the overstable photon bubble mode of Arons
(1992) as special cases.

\section{Overstability}

\subsection{Numerical Solution} 

The full dispersion relation is analytically intractable.  In the end
we will need to expand it in a small parameter to retrieve the relevant
pieces that describe the photon bubble mode.  To motivate an asymptotic
approach, we shall first solve the dispersion relation numerically.
The parameters are those appropriate to a disc around a $10 \msun$
black hole accreting at the Eddington rate, at $r = 50 G M/c^2$.
Assume $\alpha = 0.1$, so $\bm = 2.5$, and that the field is purely
vertical.  Then $\br = 0.05$, $c/c_i = 120$, $l_* = 0.07 H$, $\gamma =
5/3$, $M_0 = 0.4$.  Take $\sL = 1 - 3\br/2 = 0.93$ so that convection
is absent.  We consider a set of modes with varying $k$ and $\mu = 1/3$.
The real part of the full dispersion relation is shown in Figure 1.
The real part of the overstable mode is marked with a heavy solid line;
the imaginary part has comparable magnitude.

At large wavenumber the dispersion relation is easily interpreted
because it becomes analytic.  For general $\mu$,
\begin{eqnarray}
&(\omega^2 - {1\over{3}} c^2 k^2)(\omega^2 - {1\over{\bm}}c_i^2 k^2\mu^2) 
\times \hfill\nonumber \\
&(\omega^4 - (\br \gamma c_i^2 k^2 + {1\over{\bm}} c_i^2 k^2)\omega^2
	+ {1\over{\bm}}\br \gamma c_i^2 k^4\mu^2) \times \\
&(\omega + {i c^2\over{D_0}})^2 (\omega + {48 i (1 - \br)(\gamma - 1)
	D_0\over{\br\gamma l_*^2}}) = 0.\nonumber
\end{eqnarray}
The first term in parentheses is the flux-limited electromagnetic
wave with phase velocity $c/\sqrt{3}$ (see Mihalas \& Mihalas 1984).
The second term is the \alf wave.  The third term contains the fast
and slow MHD modes.  These are labeled in Figure 1.  Notice that the
overstable mode becomes the slow MHD mode at large wavenumber.  The final
three modes are strongly damped entropy modes.

Notice that the slow MHD mode has rather low frequency.  This is because
at short wavelengths radiation diffuses rapidly out of the perturbation
and so the radiation pressure perturbation is nil.  Thus only gas
pressure provides a restoring force for this mode.  For $\br \ll 1$,
the slow mode has $\omega^2 \approx \beta_r c_i^2 \gamma k^2 \mu^2$, so
the slow mode velocity is directly related to the sound speed associated
with the gas pressure alone.  A radiation pressure dominated fluid is
thus rather delicate on small scales in that it is easily compressed.

\subsection{Vertical Magnetic Field}

We have shown that an overstable mode exists;  we will now demonstrate the
existence of this mode analytically.  Again we consider only the simple
case of a purely vertical magnetic field.  The basic small parameter
is $\eps = (k H)^{-1}$.  The survey of conditions in accretion discs
(\S 2) then suggests the following scalings for the other parameters in
the problem.  We take $H \sim 1, c_i \sim 1, c_i/c \sim \eps, \bm \sim
1, l_* \sim 1, \br \sim \eps^{3/2}, M_0 \sim 1$, $\sL = 1 - O(\br)$,
and $\mu \sim 1$.  Using a little asymptotic foresight, we also take
$\omega \sim \eps^{-1/2}$.

Expanding $D(\omega,\bk)$ through leading order in $\eps$, we find the
remarkably simple relation
\begin{equation}\label{BUBMODE}
\omega^2 = -i g k \mu (1 - \mu^2).
\end{equation}
One root of this equation, the one with negative phase velocity, describes
the overstable photon bubble mode.

What happens if we increase the importance of gas pressure?  Suppose
that $\br \sim \eps$.  Then we find
\begin{equation}\label{SLOWDISP}
\omega^2 = -i g k \mu (1 - \mu^2) + \br c_i^2 \mu^2 k^2.
\end{equation}
The first term is the photon bubble term, while the second is that of a
slow MHD mode in which gas pressure provides the only restoring force.
If we increase $\br/(k H)$ still further, then the first term becomes
subdominant.  This branch of the dispersion relation then becomes the
slow mode, which is stable to leading order in WKB.

Since the overstability is no longer present to leading order in $\eps$
when $k \gta (\br H)^{-1}$, and $k \gta 1/H$, the overstability fails
in a WKB sense when $\br \sim 1$.  This provides an approximate 
limit on the overstable region in parameter space.

Now suppose that the diffusion time is long compared to a dynamical 
time, i.e.  $M_0 \sim \eps^{1/2}$.  Again expanding $D(\omega,\bk)$ to 
leading order in $\eps$ we find
\begin{equation}
\omega^2 + i \omega {4 c_i^2 \mu^2\over{D_0}} + i g k \mu (1 - \mu^2)
	= 0.
\end{equation}
The new term causes damping.  It shows that instability is present in
a WKB sense only on scales such that $k \gta 1/(H M_0^2)$.

\subsection{General Discussion}

We can obtain a better physical understanding of the origin of the
photon bubble instability, and a more general dispersion relation,
by expanding the linearized equations in $\eps$.   This is not trivial
because we must assign an explicit relative ordering in $\eps$ to all
the perturbed variables.   We do this by solving for the eigenvectors
$\delta f(\delta\rho)$ ($f$ represents any of the perturbed variables),
and then determining the relative ordering.  We use the same ordering
for the model parameters as that described at the beginning of \S 6.2.
This procedure reveals that the photon bubble mode consists predominantly
of motions along the magnetic field.  This is easy to understand because
the frequency of the mode, $\sim \sqrt{g k}$, is smaller by order
$\eps^{1/2}$ than the \alf frequency at the same scale, $\sim k v_A$.
Thus the field is stiff enough to resist any motion that bends the
field lines.

Writing out the dominant terms in full and allowing for a general
orientation of the initially uniform magnetic field, the continuity 
equation becomes
\begin{equation}\label{REDCONT}
-i\omega\delta\rho = -i \rho \bk \cdot \delta \bv.
\end{equation}
The perturbed velocities perpendicular to the magnetic field vanish.
Denoting the unit vector parallel to the field by $\hat{\bb}$, the parallel
component of the velocity is governed by
\begin{equation}\label{REDMOM}
-i\omega\delta \bv\cdot\hat{\bb}  = {4\pi\kappa\rho\over{c}}\delta 
	\bH\cdot\hat{\bb},
\end{equation}
so the material feels only the perturbed radiation force.
The induction equation is irrelevant, since the field is stiff.  The 
perturbed flux obeys
\begin{equation}\label{DIVH}
\bk\cdot\delta\bH = 0.
\end{equation}
The perturbed flux is conserved, so there is no exchange of energy
between the radiation field and the fluid on a timescale $\omega^{-1}$.
Combining this result with the perturbed intensity equation, we find
\begin{equation}\label{DELH}
\delta\bH = - {\delta\rho\over{\rho}}\left(
	\bH - \bk {(\bk\cdot\bH)\over{k^2}}\right).
\end{equation}
In words, the fluid feels a radiation force that is directed parallel to
the wave crests and is inversely proportional to the density perturbation.
Radiation escapes more rapidly along density minima, while energy
flow is impeded along density maxima.  Combining eqns.(\ref{REDCONT}),
(\ref{REDMOM}), and (\ref{DELH}), and denoting the unit vector along $\bk$
by $\uk$, we find the general dispersion relation
\begin{equation}\label{REDREL}
\omega^2 = -i k g (\uk\cdot\hat{\bb}) \left[ \hat{\bb}\cdot\ez - 
	(\hat{\bk}\cdot\ez) (\uk \cdot\hat{\bb})
	\right].
\end{equation}
This implies that the growth rate of the instability is largest for $k_z
= 0$ (where the WKB approximation is not strictly valid) and for $\bB$
at an angle of $\pm \pi/4$ to the wavevector.

It is readily shown that if the fluid is allowed to move freely,
unconstrained by the magnetic field, then the mode frequency vanishes
if we start from the reduced set of equations above.  More rigorously,
if we begin with the full set of linearized equations and turn off the
magnetic field, then the stability criterion reduces to the condition
that the Brunt-Vaisala frequency be real.

To understand the overstability on a more qualitative level, consider
the history of a single fluid element in the case $k_z = 0$ where $\bB$
makes an angle $\pi/4$ with the vertical.  Suppose the fluid element
initially lies in a density minimum, so it feels an increased radiation
force from below.  It is accelerated upward along the field line.
This soon puts it in a region of convergent flow and then in a density
maximum: the phase velocity of the overstable mode is such that the
density maxima progress downward along field lines.  It is now shadowed
by the density maximum, the radiation flux drops somewhat, and it falls
downward along the field line.  This puts it in a region of divergent
flow and then a density minimum, so it is accelerated upward again.
Each round of acceleration is larger than the one before, and so an
overstability results.

The photon bubble instability described by Arons is physically similar to
ours, but there are two differences.  The most important is that Arons
considers the asymptotic regime $M_o \ll 1$ appropriate to neutron star
polar caps.  This leads to a rather more complicated dispersion relation
for the overstable modes, with the growth rate proportional to $M_0^3$.
In this limit, there is a small phase offset between the radiation flux
and the density, rather than the large phase offset that initiates
the instability in the limits $M_o \sim 1$.  Another difference is
that Arons uses the Rosseland diffusion approximation, which requires
that the gas and radiation have the same temperature, while we use
a nonequilibrium diffusion approximation, which allows the gas and
radiation ``temperature'' to differ.  In the end this turns out to have
no effect whatsoever on the overstable modes, because gas pressure (and
hence gas temperature) is completely negligible for modes with frequency
and wavelength comparable to the photon bubble mode.  Finally, we note
that in the appropriate limit we are able to recover Arons's dispersion
relation, thus confirming his analysis.

\section{Nonlinear Outcome}

The astrophysical implications of the photon bubble instability depend on
the nonlinear outcome, to which the linear theory is an unreliable guide.
Numerical experiments by Hsu et al. (1997) show that in the regime $M_0
= c/(\tau_{es} c_i) \ll 1$, $\bm \ll 1$ the photon bubble instability
leads to greatly enhanced vertical transport of energy.  It thus seems
likely that the photon bubble instability will enhance vertical energy
transport in discs.

It is tempting to make an analogy between photon bubbles and convection.
Convection generates such efficient transport of energy that it erases the
inverted entropy gradient that initiated it and, if forced, maintains the
convective fluid in a marginally stable state.  It is natural to think
that the photon bubble instability drives the disc toward a marginally
stable state as well.  Our analysis does not give any rigorous stability
criteria, but it does show that when $\beta_r \sim 1$ the instability
is no longer present at leading order.  Thus the instability is likely
to be absent or greatly reduced in strength when $\beta_r \sim 1$.
If the disc is initially radiation dominated, photon bubbles might then
transport energy efficiently out of the disc, lowering the temperature
and raising $\br$ toward $1$.

The nonlinear outcome of the photon bubble instability can be
characterized by a cooling rate $Q^-(\Sigma,T_c)$, which is the energy
lost per unit time per unit disc area.  The thermal and viscous stability
properties of the disc depend on how $Q^-$ varies with temperature and
surface density (Piran 1978).  Ultimately this can only be evaluated
from a fully nonlinear theory, but if $Q^-$ is a steep enough function
of $T_c$, then the disc can be viscously and thermally stable.  

Our analysis also shows that when the dimensionless radiative diffusion
rate $M_0$ decreases the instability weakens, in that instability is
only present for $k H \gta M_0^{-2}$.  This effect may also shut
off photon bubbles.  At low accretion rates, for example, a disc with
$\beta_r \sim 1$ will have large $M_0$, while at larger accretion rates
$M_0$ is smaller.

Photon bubbles might also change the emergent spectrum.  The bubbles
make the disc porous to radiation, so a photon traverses a shorter
path to the surface than it would if the disc were subject only to
ordinary radiative difffusion.  The photon distribution therefore has
less opportunity to thermalize.  In addition, the vertical transport
might be more episodic, enhancing variability.

This discussion is speculative.  Other physical processes may contribute
to vertical energy transport in discs: ordinary convection (but see
the discussion of Rees 1987) or magnetic Rayleigh-Taylor instability
(see the flux-tube calculation of Sakimoto \& Coroniti 1989) may
dominate photon bubbles.  In addition, MHD turbulence initiated by the
Balbus-Hawley instability must coexist with photon bubbles.  We have
developed a linear theory only when these effects are absent.  The full
nonlinear development of the radiation dominated disc can probably only
be studied realistically via three dimensional numerical experiments.
Numerical methods exist for treating the radiative transfer in a
flux-limited Rosseland diffusion approximation.  Since the character
of the unstable mode is identical in the nonequilibrium and Rosseland
diffusion approximations, numerical studies of the nonlinear evolution
of a radiation dominated disc may be immediately practical.

\section{Summary}

We have considered the linear theory of a radiation dominated atmosphere
with spatially constant magnetic field as a model for the radiation
dominated inner parts of thin accretion discs around compact objects.
The model is subject to an overstable photon bubble mode that tends
to separate radiation and matter.  The photon bubble dispersion
relation for a general orientation of the magnetic field is given by
eqn.(\ref{REDREL}).  Vertical energy transport is likely to be enhanced
in the nonlinear outcome.  The disc may then deflate until it is no
longer radiation pressure dominated.

I thank J. Krolik, R. Narayan, M. Rees, G. Rybicki, E. Vishniac,
and the referee, E. Szuszkiewicz, for their discussions and comments.
This research was supported in part by NASA grant NAGW 5-2837.

\section*{Figure Captions}

\begin{figure}
\centerline{\epsfig{figure=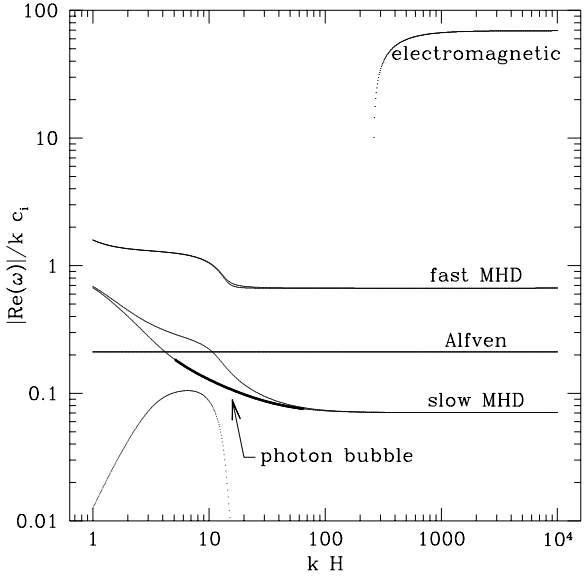, width=8.1cm}}
\caption{
Figure 1.  The real part of the WKB dispersion relation for a
radiation-dominated atmosphere with uniform vertical field.  See text
for details.  The abscissa is the wavenumber in units of the scale
height; the ordinate is the phase velocity in units of the isothermal
sound speed $(P/\rho)^{1/2}$.  The real part of the phase velocity for
the photon bubble mode is shown as a heavy line.  The growth rate
(imaginary part) is comparable in magnitude.
}
\end{figure}

\end{document}